\documentstyle[aps,prd,preprint,tighten,floats,epsf]{revtex}

\begin{document}

\preprint{VAND-TH-98-19
\hspace{-30.0mm}\raisebox{-2.4ex}{November 1998}}

\title{The Interaction Structure and
Cosmological Relevance of Mass Scales in String Motivated Supersymmetric
Theories} 

\author{Arjun Berera\thanks{E-mail:
bereraa@ctrvax.vanderbilt.edu} and
Thomas W. Kephart \thanks{E-mail: kephartt@ctrvax.vanderbilt.edu}}

\address{
Department of Physics and Astronomy, Vanderbilt
University,
Nashville, TN 37235, USA}

\maketitle

\begin{abstract}

A string motivated, N=1 global supersymmetric, general Lagrangian is
obtained that represents an inflaton interacting with the modes of a
string.  {}Focus is placed on the Lagrangian's relevance to warm
inflation, which is a cosmology based on dissipative dynamics.
Several interaction schemes are identified  from the general Lagrangian
that may have interesting consequences for dissipative dynamics.
Distributed-mass-models, which have been shown to solve the horizon
problem within a warm inflation regime, are identified in the general
Lagrangian.

\vspace{0.34cm}

\end{abstract}

\medskip

hep-ph/9811295

\bigskip

In Press Physics Letters B 1999

\eject

\section{Introduction}

Warm inflation cosmology \cite{wi,ab1}
has incited a detailed study of the
nature of interactions in particle physics models at
energies from the electroweak scale to the Planck scale.
Warm inflation is comprised of non-isentropic expansion in the
background cosmology \cite{rudnei,gmn,ab2} and thermal seeds of density
perturbations \cite{bf2}.  Within the warm inflation regime, while the
scale factor undergoes inflationary expansion, the radiation energy
density is still substantial due to its constant production from
conversion of vacuum energy.  This expansion regime is intrinsically
different from the supercooled inflation regime, since in the warm
inflation case the inflationary period smoothly terminates into a
subsequent radiation dominated regime, without a reheating period.

The simplest field theory description of the
non-isentropic background cosmology is from Ginzburg-Landau overdamped
relaxational kinetics \cite{ab2}.  
The quasi-adiabatic regime of such kinetics
implies it can be systematically computed in quantum field theory
\cite{hs1,morikawa,lawrie,gr,yy,bgr,yl} and the overdamped limit has been
explicitly demonstrated \cite{bgr}.  Such kinetics, in a cosmological
warm inflation setting requires several additional and nontrivial
consistency conditions between the macroscopic and microscopic dynamics.
In a specific quantum field theory model which has been named the
distributed-mass-model (DM-model) \cite{bgr,abpas}, it recently has been
demonstrated \cite{bgr2} that all the consistency conditions can be satisfied
with sufficient overdamping to solve the cosmological horizon problem.
These models are atypical to particle physics due to a certain shifted
interaction scheme which is detailed in Sect. III.
Furthermore, the relevant DM models for the warm inflation 
results in \cite{bgr} require a large number of fields, $> 10^4$.

The success of DM models to warm inflation
cosmology immediately prompts questions about their
fundamental origin and more generally
exemplifies the importance of the specific interaction structure in the
underlying particle physics model for the eventual realizability of a
first principles warm inflation cosmology.  In \cite{abpas,bgr2}
it was mentioned that DM-models have heuristic similarity to
string theory, but no explanation was offered.  In general, for the
energy scales of interest to cosmology, the obvious place to examine the
variety of particle physics interaction schemes 
is effective supersymmetric theories
derived from strings.  This is a vast 
subject for which an {\it ab initio}
approach at this moment would be impossible.  As a first step, this
paper attempts an interpretation of effective low energy 
SUSY theory motivated from string
theory that is conducive to realizing warm inflation cosmology.
The specific example of DM-models will be reviewed and
we will demonstrate their origin
through this interpretation.  En route we
will also note the variety of possible other
particle physics models which may be relevant to warm inflation cosmology.

\section{Motivation from String Theory}

We will assume warm inflation takes place 
somewhere below the Planck scale $M_{Pl}
\sim 10^{19} {\rm GeV}$ but above some 
${\rm GUT}$ scale $M_{\rm GUT} \sim 10^{15}
{\rm GeV}$. In string theory the region between these two scales is not
devoid of states.  Quite the contrary, the generic string scale is
$M_{S} = 5.3 g_S \times 10^{17} {\rm GeV}$ \cite{Kaplunovsky} 
so there can be
$O(10^2)$ mass levels between $M_{\rm GUT}$ and $M_{Pl}$.
{}Furthermore, as the mass level increases, the number of states per level
grows rapidly \cite{loweng}.
The states at each mass level fall into representations of the symmetry
of the theory, gauge, global and Lorentz \cite{gsw}.
As the temperature falls, these symmetries can change. Let us follow the
gauge symmetry. The gauge group $G$ may break to some smaller group $G'$
as the temperature falls. Each G-irrep $R$ at the N-th level of the
theory will decompose into irreps that are a sum of $G'$ irreps , $R'$,
i.e., $R \rightarrow \sum_i R_i'$, when the symmetry is broken.
This creates a fine structure at each energy level. Similar arguments
hold for global symmetry breaking and for reduction of the Lorentz
symmetry via compactification.

In the next section, the single aspect of string theory we need
is that there are many mass levels.  To our knowledge, the
effective theory in the next section is the first to make use of the higher
levels of strings for phenomenological purposes. 
As such, we will focus only on the basic point and
leave detailed model building for future work.

Between gauge symmetry breaking, SUSY breaking and compactification,
there are many fields that could be considered as potential inflatons.
Several of these possibilities have been recently reviewed (see
\cite{lyth}). A few fields of particular interest are singlet
scalar fields involved in the {}Fayet-Iliopoulos SUSY breaking
mechanism, singlets involved in {}F-term spontaneous SUSY breaking via
the O'Raifeartaigh mechanism, gauge singlet superfields
in flat direction of the superpotential, possible singlet winding
modes associated with Wilson loop gauge symmetry breaking in the
string theory, etc., not to mention the universal favorite, the
dilaton. Rather than make a choice, here is will be assumed that we have
started with a fundamental string theory, and we then study only the
effective SUSY gauge theory below the compactification scale, but
above $M_{GUT}$, where an inflaton has already emerged at a high field
value along with a tower of states in representations of the gauge
group, all weakly coupled to the inflaton.  There can be many fields that
do not mix with the inflaton. As these are of no interest for the
following discussion, terms involving them will not be displayed.
As the inflaton descends
through the tower of states causing dissipation, it engenders the
warm inflation scenario.

Before writing down a specific SUSY model with the desired warm
inflation
properties, two remarks about supersymmetry are in order
({}For a complete review of supersymmetry, please see \cite{nilles,clark}.).
{}First, as long as global supersymmetry is a good symmetry, {}F-terms are not
renormalized (meaning that once we choose the parameters in the
superpotential, they stay fixed until SUSY is broken).  Because of
this nonrenormalization theorem \cite{gsr}, the fixing of
{}F-term parameters is natural in the technical sense, and for that
reason has been invoked to alleviate difficulties like the hierarchy
problem. (D-terms in general are renormalized and can
therefore run with energy scale, but this is not
always true \cite{fischler}.) Amongst other things, the nonrunning
of the {}F-terms allows one to fix flat directions in the
superpotential.  The second comment about SUSY is the 
SUSY Hamiltonian, $H$, can be
written as the sum of the squares of the supercharges $Q_{\alpha}$.
SUSY is unbroken if the vacuum $\Omega$ is invariant under
SUSY transformations, $Q_{\alpha}| \Omega \rangle=0$. 
In the case that SUSY is unbroken, it therefore
implies the vacuum energy is zero, 
$\langle H \rangle_{\Omega} \equiv \langle \Omega | H | \Omega
\rangle=0$.  As such, a necessary and sufficient condition for SUSY to
be spontaneously broken is a 
nonvanishing vacuum energy $\langle H \rangle_{\Omega}>0$
\cite{witten}.

Since we plan to deal only with the effective global SUSY theory, we
will not need to make a specific choice of where the inflaton lives
within the overlying string theory but as we have noted, many
possibilities for its origin exist. Even after SUSY breaking by
{}F-terms, mass sum rules \cite{fgp} exist and
constrain the running of masses. This provides useful knowledge as the
gauge group $G$ breaks and the states rearrange into representations
of $G'$. Although in general the sum rules are altered by D-term SUSY
breaking, under certain circumstances anomaly cancellation requires
their preservation.

\section{Effective Lagrangian}

Let us consider the general form of an effective N=1 global SUSY theory
representing an inflaton interacting with the modes of a string.  Such a
theory consists of a single chiral superfield $\Phi$ which represents
the inflaton and a set of chiral superfields $X_i$, $i=1,\ldots,N_M$
representing the string modes.  All the superfields have their
antichiral superfields $\bar \Phi$, $\{\bar X_i\}$ appearing in kinetic
and Hermetian conjugate (h.c.) terms.  In the
chiral representation the expansion of the superfields in terms of the
Grassmann variable $\theta$ is
$\Phi=\phi + \theta \psi + \theta^2 F$ and 
$X_i= \chi_i + \theta \psi_i + \theta^2 F_i$, $i=1, \ldots , N_M$.
Here $\phi=(\phi_1+ i \phi_2)/\sqrt{2}$ and
$\chi_i = (\chi_1+ i \chi_2)/\sqrt{2}$ are complex scalar fields,
$\psi$ and $\{ \psi_i \}$ are Weyl
spinors, and $F$ and $\{F_i\}$ are auxiliary fields.   
By definition, the inflaton $\Phi$ characterizes the state
of the vacuum energy through a nonzero amplitude in the bosonic sector
$\langle \phi \rangle \equiv \varphi_c \neq 0$.
Warm inflation cosmology specifically recognizes that the interaction of
the inflaton with the other fields can lead to nontrivial dissipative
effects that both convert vacuum energy into radiation energy and
slow the motion of $\varphi_c$.

The general Lagrangian for this system of fields is composed of a
kinetic and interacting part 
\begin{equation}
{\rm L} = {\rm L}_{K} + {\rm L}_{I}.
\label{genl}
\end{equation}
The kinetic energy portion of the Lagrangian is
\begin{equation}
{\rm L}_{K} = \int d^4x d^2 \theta d^2 {\bar \theta}
\left[ {\Phi}{\bar \Phi} + \sum_{i=1}^{N_M} X_i {\bar X}_i
\right] ,
\label{lkin}
\end{equation}
where the fields $F$,$\{F_i\}$ do not appear, thus in fact are auxiliary
fields.  The interaction Lagrangian is
\begin{equation}
{\rm L}_I = \int d^4x d^2 \theta  W(\Phi,\{X_i\}) +
{\rm h.c.} + {\rm interaction \ D-terms}.
\label{li}
\end{equation}
Considering only {}F-terms, the general superpotential for this system
is 
\begin{equation}
W(\Phi,\{X_i\})= 4m {\Phi}^2 + \lambda \Phi^3
+\sum_{i=1}^{N_M} \left[4\mu_i X_i^2 + f_i X_i^3 +
\lambda'_i \Phi^2 X_i + \lambda^{''}_i \Phi X_i^2 \right].
\label{superpot}
\end{equation}
If $\{X_i\}$ represent the modes of a string, then the mass parameters
$\{\mu_i\}$ will range over all energy levels of the string
from the string scale $M_S$ to the
Planck scale $M_{Pl}$ and beyond.  
The energy levels can represent both the coarse
splittings between mass levels of the symmetry unbroken theory and the
fine structure splitting after some symmetry breaking. Here we will not
commit to a specific model.
The auxiliary fields $F$,$\{F_i\}$ can be eliminated in
Eqs. (\ref{lkin}), (\ref{li}) and (\ref{superpot}) 
through the ``field equations''
$\partial W/\partial F = \partial W/\partial F_i =0$, with the resulting
Lagrangian only in terms of the physical fields 
$\phi$, $\psi$, and $\{\chi_i,\psi_i\}$.

\subsection{Distributed Mass Model}

The general Lagrangian involves several interesting interaction schemes
between the inflaton and the mode-fields. As an example,
the DM-models will be obtained from the general Lagrangian
Eqs. (\ref{genl}) $-$ (\ref{superpot}).  {}First, let us take
a moment to review the DM-model;
these are models where the fields $\{\chi_i,\psi_i\}$ that couple
to the inflaton $\phi$, have masses that are dependent on the inflaton's
amplitude $\varphi_c$ and for any given $\varphi_c$ there is a wide
range of mass scales for the system of fields $\{\chi_i,\psi_i\}$.
Such models are obtained by the shifted couplings 
$g^2 (\phi-M_i)^2 \chi_i^2$, $g (\phi-M_i) {\bar \psi_i}{\psi_i}$
(for Weyl fermions mass terms are of the form
$g (\phi-M_i) \psi_i \psi_i$ and $g (\phi-M_i) {\bar \psi}_i {\bar \psi}_i$)
for bosons $\chi_i$ and fermions $\psi_i$ respectively, with
$\{M_i\}$ ranging over mass scales that can be both coarsely and finely
split.
The important reference
energy scale during warm inflation is the temperature T of the universe.
When the mass of a specific $\chi_i$($\psi_i$) field
$m_{\chi_i(\psi_i)} \approx g |\varphi_c-M_i| \stackrel{<}{\sim} T$, 
that field is thermally excited and can
contribute to the dissipative dynamics of the inflaton, but outside this
region it cannot \cite{bgr,bgr2}.
By having many mass scales $\{M_i\}$, the 
$\varphi_c$-amplitude region for active 
dissipative dynamics is increased.
This directly translates to an increased duration of warm inflation,
thus increasing the total e-foldings.  

The DM-models are contained in the class of Lagrangians 
from Eqs. (\ref{genl}) $-$ (\ref{superpot}) in which
$\lambda_i^{\prime}=0$, $i=1, \ldots, N_M$.  {}For this case
writing $L=\int d^4x {\cal L}$, we obtain
\begin{equation}
{\cal L}={\cal L}_{\Phi} + \sum_{i=1}^{M} {\cal L}_{X_i-{\rm kinetic}}
+\sum_{i=1}^M {\cal L}_{X_i-{\rm mass}} +
{\cal L}_{I}
\end{equation}
where
\begin{eqnarray}
{\cal L}_{\Phi} &=& \partial_\alpha \phi^{\dagger} \partial^\alpha \phi
-4m^2 \phi^{\dagger} \phi  \nonumber \\
&-&\frac{3m\lambda}{2} (\phi^{\dagger} \phi^2 + \phi^{\dagger 2}\phi)
-\frac{9}{16} \lambda^2 (\phi^{\dagger} \phi )^2 \nonumber \\
&+&\frac{i}{4} \psi \stackrel{\longleftrightarrow}{\not\!\partial} {\bar \psi}
+\frac{m}{2}(\psi \psi+ {\bar \psi} {\bar \psi})
+\frac{3}{8} \lambda (\phi \psi \psi + 
\phi^{\dagger} {\bar \psi} {\bar \psi}),
\label{ipot}
\end{eqnarray}
\begin{eqnarray}
{\cal L}_{X_i-{\rm kinetic}}=
\partial_\alpha \chi^{\dagger}_i \partial^\alpha \chi_i
&+&\frac{i}{4} \psi_i 
\stackrel{\longleftrightarrow}{\not\!\partial} {\bar \psi}_i,
\end{eqnarray}
\begin{eqnarray}
{\cal L}_{X_i-{\rm mass}}&=& -(2\mu_i + \frac{\lambda_i^{''}}{2}
\phi^{\dagger})(2 \mu_i + \frac{\lambda_i^{''}}{2} \phi)
\chi^{\dagger}_i \chi_i \nonumber \\
&-&(\frac{\lambda_i^{''} m}{2} \phi^{\dagger} +
\frac{3\lambda \lambda_i^{''}}{16} \phi^{\dagger 2}) \chi_i^2 
-(\frac{\lambda_i^{''} m}{2} \phi+
\frac{3\lambda \lambda_i^{''}}{16} \phi^2) \chi_i^{\dagger 2} 
\nonumber \\
&+&[\frac{\mu_i}{2} +\frac{\lambda_i^{''}}{8} \phi] \psi_i \psi_i
+[\frac{\mu_i}{2} +\frac{\lambda_i^{''}}{8} \phi^{\dagger}] 
{\bar \psi_i}{\bar \psi_i},
\end{eqnarray}
and
\begin{eqnarray}
{\cal L}_{I}&=& \sum_{i=1}^{N_M} [
\frac{\lambda_i^{''}}{4}(\psi \psi_i \chi_i+
{\bar \psi}{\bar \psi_i} \chi_i^{\dagger})
+\frac{3f_i}{16}(\chi_i \psi_i \psi_i + 
\chi_i^{\dagger} {\bar \psi_i} {\bar \psi_i} ) \nonumber \\
&-& \frac{3}{2} \mu_i f_i (\chi_i^{\dagger 2} \chi_i+
\chi_i^{\dagger} \chi_i^2)
-\frac{9}{16} f_i^2 \chi_i^{\dagger 2} \chi_i^2
-\frac{3f_i \lambda_i^{''}}{16}
(\phi^{\dagger}\chi_i^{\dagger}\chi_i^2+\phi \chi_i \chi_i^{\dagger 2})
] \nonumber \\
&-& \frac{1}{16} \sum_{ii'}^{N_M} 
\lambda_i^{''}\lambda_{i'}^{''} \chi_i^{\dagger 2} \chi_{i'}^2.
\label{lint}
\end{eqnarray}

${\cal L}_{X_i-{\rm mass}}$ contains all the terms that contribute to the mass
of the $\chi_i,\psi_i$ fields for arbitrary inflaton amplitude
$\varphi_c$.  The DM-model is realized for the case
$\mu_i=gM_i/2$ and $\lambda_i^{''}=-2g$.  {}For this case the masses of the
$\chi_i,\psi_i$ fields are respectively
\begin{equation}
m^2_{\chi_i}= g^2(\varphi_c-M_i)^2 -2gm \varphi_c -
\frac{3g \lambda}{4} \varphi_c^2 
\label{mchi}
\end{equation}
and
\begin{equation}
m^2_{\psi_i}= g^2(\varphi_c-M_i)^2 .
\label{mpsi}
\end{equation}
At $\varphi_c =0$, the masses of the $\chi_i,\psi_i$ pair are
equal in Eqs. (\ref{mchi})
and (\ref{mpsi}) as required by supersymmetry.  On the other hand, a nonzero
inflaton field amplitude, $\varphi_c \neq 0$, implies a breaking of
supersymmetry, which in turn permits the differences in the
$\chi_i$ and $\psi_i$ masses in Eqs. (\ref{mchi}) and (\ref{mpsi})
respectively.  In this case, $\varphi_c$ is out of equilibrium,
thus SUSY breaking here is not due to one of the standard equilibrium
processes,
for example O' Raifeartaigh \cite{oraf}.  However for this case, when the field
amplitude comes into equilibrium, $\varphi_c=0$, SUSY is restored.
These breaking terms could be potentially dangerous
to the bosonic sector of the mode fields, 
if their size is much larger than the temperature
scale T.  In such a case, these bosonic mode fields would be too massive to
contribute to the dissipative dynamics of $\varphi_c$.
However, this problem will not arise for the wide set
of cases obtained in
\cite{bgr2}, in which the parameter regime required
$m=m_T \sim \sqrt{\lambda} T$ and 
$\lambda \varphi_c^2 = m_{\phi}^2 < T^2$. {}For these values of the
parameters, from inspection of Eq. (\ref{mchi}), it follows that the
soft breaking terms are smaller than $T^2$.

Based on the analysis in \cite{bgr2}, the amplitude of the inflaton
$\varphi_c$ and mass parameters $\{\mu_i\}$ 
are expected to be generically much larger than
T.  In this case, terms in the interaction Lagrangian Eq. (\ref{lint})
involving $\varphi_c$ and $\{\mu_i\}$ could be very large and
lead to interesting dissipative effects.  This deserves further
investigation, but it will not be done here.  It is worth adding that
the interaction terms in Eq. (\ref{lint})
involving $\varphi_c$ and $\mu_i$ couple through the
parameter $f_i$.  Thus the effect of these terms can be controlled and
if desired they can be made negligible by choosing $f_i$ sufficiently
small.

\section{Conclusion}

In the previous section we obtained the DM-models from the general SUSY
Lagrangian Eqs. (\ref{genl}) $-$ (\ref{superpot}).  
Warm inflation cosmology based on DM-models
recently has been shown to solve the horizon problem \cite{bgr2}.
Thus the results in the previous section highlight a specific class of
potential SUSY warm inflation models. 

The general SUSY Lagrangian Eqs. (\ref{genl}) $-$ (\ref{superpot}) 
contains other useful features
for warm inflation cosmology.  {}For example, the inflaton potential
in Eq. (\ref{ipot}) has cubic $\phi^3$ terms which 
can produce plateau
regions that have significant vacuum energy and reasonably flat
potential surfaces.  The dual effect of a flat potential and strong
dissipation increases the duration of warm inflation, which is being
driven by the vacuum energy that the plateau maintains.
Another interesting feature of the SUSY 
Lagrangian Eqs. (\ref{genl}) $-$
(\ref{superpot})
is for $\lambda'_i \neq 0$ there are $\phi^3 \chi_i$ couplings
in which the bosonic mode fields $\chi_i$ couple linearly to the inflaton.
Such terms could be treated nonperturbatively in deriving the effective
$\varphi_c$-equation of motion using methods similar to
those for solving
Caldeira-Leggett models \cite{cl,ab1}.
An exact nonperturbative derivation of the
effective $\varphi_c$-equation of motion would not be possible, since
there still would be additional terms in the Lagrangian that would
require a perturbative treatment.
Nevertheless, such interactions provide a novel prospect for dissipative
dynamics. {}Finally in regards the general SUSY Lagrangian
Eqs. (\ref{genl}) $-$ (\ref{superpot}), D-terms generically are products of
chiral and antichiral fields, such as the kinetic terms 
eq. (\ref{lkin}) and interaction terms from a gauge symmetry.
Interaction D-terms are not required to obtain the
DM-model mass scale structure. However, where present, they modify the
Higgs potential, adding robustness to the model but at the same time
constraining its parameters.  {}For dissipative dynamics, the enhanced
interactions amongst the mode fields due to 
interaction D-terms will increase the mode field
decay widths, thus thermalization rates \cite{hs1,lawrie,gr,bgr}.

In summary, we have argued that DM-models arise when we
consider effective N=1 SUSY Lagrangians obtained from a fundamental
string.  For observationally interesting warm inflation 
solutions in \cite{bgr}, the relevant DM models require a large number of
fields, $> 10^4$. 
The tower of states between the string scale and Planck scale
are made to order for such a realization of warm inflation.
This is virtually independent of the field that behaves as the inflaton.
All we require is the existence of such a state.  A more complete
analysis would derive these results from string theory, but this is
currently out of reach and beyond the modest objectives of this work
as well.

\section*{Acknowledgments}
We thank Marcelo Gleiser, Rich Holman and 
Rudnei Ramos for helpful discussions.
This work was supported by the U. S. Department of Energy
under grant DE-FG05-85ER40226.

\end{document}